\magnification1200

\rightline{KCL-MTH-00-4}
\rightline{hep-th//0001216}

\vskip .5cm
\centerline
{\bf{{Automorphisms, Non-linear Realizations  and  Branes }}}
\vskip 1cm
\centerline{ P. West}
\vskip .2cm
\centerline{Department of Mathematics}
\centerline{King's College, London, UK}

\leftline{\sl Abstract}
The theory of non-linear realizations is used to derive the dynamics
of the branes of M theory. A crucial step in this procedure is to use
the enlarged automorphism group of the supersymmetry algebra
recently introduced. The field strengths of the
worldvolume gauge fields arise as some of the Goldstone fields
associated with this automorphism group. The relationship  to the
superembedding approach is given.
\vskip .5cm

\vfill\eject


\parskip=18pt

\medskip
{\bf {0. Introduction }}
\medskip
Many  superbranes were constructed using a
generalisation of the  $\kappa$-symmetry [1] first found in the point
particle and later in the string. These included the construction of
the action for  the twobrane of M theory [2] and the D-branes in ten
dimension [3]. However, the construction of the dynamics of the most
sophisticated brane, the M theory fivebrane, was  achieved
[4],[5],[6] using   the superembedding formalism [7]. In fact, some
the first brane actions to be constructed were for  superbranes in
six and  two dimensions [8] using  the method of
non-linear realizations [9],[10]. This method has some advantages,
namely the construction is entirely group
theoretical and the fields have a very natural interpretation in terms
of Goldstone particles for broken symmetries. There now exists a
substantial literature [11][12][13][20] on the construction of
superbrane actions using this method. However, these works are
 largely confined to branes in
dimensions lower than ten dimensions and   the branes in eleven and
ten dimensions have not been constructed using this method.
\par
While the relationship between the
superembedding approach and that using  $\kappa$-symmetry is well
understood in the sense that the  $\kappa$-symmetry arises as part
of the worldvolume supersymmetry which is automatically present in
the superembedding approach, the relationship between these
approaches and the approach of non-linear realizations is not clear
 although some discussion is given in reference [14].
\par
It is the purpose of this paper to give a
systematic method of deriving the  dynamics  of some of the more
commonly used superbranes using the theory of
non-linear realizations. A crucial step in this construction
is to include in the group not only the usual Spin group, but its
generalisation to a much larger automorphism that exists for
 supersymmetry algebras whose     central charges form an
arbitrary matrix. This group was introduced  in reference
[15] and applied within the context of  the symmetries of  the
covariant equations of motion of  the M theory fivebrane. In
particular, it was shown that the currents for supersymmetry
translations and the second rank central charge arose by expressing
the equations of motion for the fermion, scalars and second rank
tensor gauge field respectively as total derivatives. This suggests
that even this brane possessed a construction as a non-linear
realization. It was also shown [15] that the fivebrane
possessed a further symmetry with a rank three generator
that was part of the $GL(32)$
automorphism group of the eleven dimensional supersymmetry algebra with
all its central charges. The extra generator  was associated
with the two form gauge field of the fivebrane.
These  large automorphism groups are only present if one includes
sufficient central charges in the supersymmetry algebra. Indeed, if
one includes all the possible central charges  in the
superalgebra, then  the
automorphism group is $GL(c_d)$
 where $c_d$ are the number of Majorana supercharges. This is the
natural generalisation  to branes of the  Spin
 and  internal symmetry groups that occur if one considers
only point particles and so has only a momentum generator in the
anti-commutator of the supercharges. In  the past, many constructions
of branes  have not taken into
account in the coset construction any
automorphisms of the supersymmetry algebra.   Some notable exceptions that
did include the well known Lorentz and internal groups were given in
references [11, 20, 24] which treated some
superbranes in six and four  dimensions respectively and the 
point particle in lower dimensions respectively.
As we shall
see,  to construct  the branes of M theory using the method of
non-linear realisations one must use the much larger automorphism
groups. This is consistent with the old result that the
topological charges of branes occur as  the central charges in the
supersymmetry algebra [23].
In the approach of non-linear realizations
given in this paper,
the worldvolume fields of the superbrane are certain of the Goldstone
fields associated with this automorphism group.
\par
We will also find the
relationship between the superembedding approach and that of the
non-linear realizations. In the latter one first derives  the Cartan forms
from the group elements  and then  constructs from them a
 group invariant dynamics. It was noticed long ago that for
certain choices of groups
one could set certain of the Cartan forms to vanish in a group
invariant manner. This was the so-called inverse Higgs effect [16].
We will show that for the supergroups we consider, we can choose
 such a constraint on the Cartan forms which is
precisely equivalent to the condition  assumed in the superembedding
formalism.

\medskip
{\bf {1. The Theory of Non-linear Realizations }}
\medskip
We first briefly review the  theory of non-linear
realizations set out by Coleman, Wess and Zumino [9] and extended
by Volkov [10]. Rather than consider a general group we restrict
ourselves to the   (super) group
$\underline G$ whose generators can be divided into the two sets
$\underline K$ and $\underline H$. The sets $\underline K$ and
$\underline H$ are both supgroups of $\underline G$ and
$\underline H$ is the automorphism group, of $\underline K$. 
 The generators in each of the above two classes    are further divided as
$\underline K=\{ K, K'\}$ and
$\underline R=\{R, R'\}$.  The generators $K$ and $R$ both form
subalgebras of  the  Lie
(super) algebra $G$ whose associated groups
 we denote by $K$ and $H$ respectively.
The generators of $H$ are an automorphisms  of $K$.
The division of the generators
of
$\underline G$ into these four classes corresponds to:
\medskip
\item { $K$}, unbroken generators associated with positions in
(super) space-time,
\item { $K'$}, spontaneously broken generators associated with
positions in (super) space-time,
\item { $R$}, unbroken automorphism generators,
\item { $R'$}, spontaneously broken auotmorphism generators.
\medskip
The automorphism generators include those of the Lorentz group or its
covering  group the corresponding Spin group,  
in some cases  internal group
generators,   but also other generators which are not usually
considered.
\par
>From now on we will drop the prefix (super), but  it is to be
understood to be present.  As will be clear, the objects associated
with the group $\underline G$ carry an underlined indices while those
associated with the subgroup $G$ carry no underline.  The
decomposition of the former into the latter and the remainder
 is achieved using unprimed and primed indices respectively.
\par
We wish to consider the coset
$\underline G/H$.   For simplicity we will use an exponential
description of group elements and we may then choose
 the coset representatives to be 
$$g= e^{X\cdot K}e^{X'\cdot K'}e^{\phi'\cdot R'}
\eqno(1.1)$$
In this equation $\cdot$ denotes the relavent summation over the
indices. Under any rigid group transformation
$g_0$ we find that
$$g\to\ g_0 g= \hat g= e^{\hat X\cdot K}e^{\hat X'\cdot
K'}e^{\hat \phi'\cdot R'} h_o
\eqno(1.2)$$
where $h_0$ is an element of $H$. The Cartan forms are given by
$$g^{-1}dg= F\cdot K + F' \cdot K' + \omega '\cdot R'
+\omega \cdot R
\eqno(1.3)$$
Under equation (1.2) the Cartan forms transform as
$$\hat g ^{-1}d\hat g=  h_0(g^{-1}dg )h^{-1}_0 +h_0 d h^{-1}_0
\eqno(1.4)$$
It can happen that the Cartan forms carry a reducible representation
of $H$, in which case, certain of the forms can be set to zero. This is
the so called inverse Higgs effect [16]. It has the effect of
eliminating some of the Goldstone fields  in terms  of the
others.  The action or equations of motion are to be constructed  from
the Cartan forms   in such a way  that they are
invariant under the transformations of equation (1.2).
\par
The conventional interpretation of the above equations is to
regard the $X$ as the coordinates of (super) space-time and
to take the fields $X'$ and $\phi'$ to depend on them. This leads to a
field theory on the coset space $G/H$.
This approach has been almost universally adopted.  However, when
considering branes it is more instructive to consider a more general
possibility. The brane is moving through the coset space $\underline
G/ \underline H$ with tangent group $\underline H $ and sweeps out a
submanifold that has the dimensions of the coset $G/H$ and a tangent
space group $H$.  We therefore
consider the representatives of the coset $\underline G /H$ of
equation (1.2) to depend on the  variables $\xi$ which parametrises the
embedded  submanifold.  Since the Cartan forms involve the exterior
derivative
$d$ they are independent of the coordinate system used.
The Cartan forms associated with $K$ are given by
$F \cdot K= d\xi\cdot F\cdot K$ and similarly for the other Cartan
forms. We can think of the $F$ in this equation as the vielbein on
the embedded submanifold. The covariant derivatives of the
Goldstone fields associated with $K'$ are defined by
$$F' \cdot K'\equiv F\cdot\Delta X'\cdot K' = d\xi\cdot F^{-1}\cdot
F'\cdot K'.
\eqno(1.5)$$
The $\Delta X'$  are independent of reparamenterisations.
A similar construction can be made for the covariant
derivatives of $\phi'$. When identifying the fields that can be
set to zero, i.e. use
the inverse Higgs mechanism,  we must  not only maintain the
$\underline G$ invariance of equation (1.2) and also
reparameterisation invariance.   In effect, this means setting only
those covariant derivatives of the Goldstone fields in equation (1.5) 
that transform
in a covariant manner under $h_0$ to zero.
The equations of motion,  or action,  are to be constructed from the
vielbein and the covariant derivatives of the Goldstone fields that
remain. In this way one can find a formulation of brane dynamics
that is reparameterisation invariant and also invariant under the
rigid $\underline G$ transformations of equation (1.2). From this
approach we can recover the more  conventional approach  by using the
reparameterisation invariance to choose static gauge, i.e.
$X=\xi$ for those coordinates that lie in the brane directions.
\medskip
{\bf {2. Bosonic Branes }}
\medskip
 In this section we will show that the dynamics of bosonic p-branes in
a flat background in $D$ dimensional space-time arises as a
non-linear realization in the sense of the previous section. We
take $\underline G=ISO(D-1,1)$ with $\underline  K =
\{ P_{\underline n }\} $ and $\underline H= \{
J_{\underline n \underline m}\}$ where $\underline n, \underline
m=0,1\ldots, D-1$ and
$G=ISO(p,1)$ with $K=\{P_n\}$ and $H= \{ J_{nm}\}$
where $n,m= 0,1,\ldots ,p+1$.
This is to be expected as the presence of the p-brane clearly breaks
the background space-time group $ISO(D-1,1)$ to $ISO(p,1)$. The Lie
algebra of $ISO(D-1,1)$ is given by
$$[J_{\underline n \underline m},J_{\underline p \underline q}]
=-\eta_{\underline n \underline p}J_{\underline m \underline q}
-\eta_{\underline m \underline q}J_{\underline n \underline
p}+\eta_{\underline n \underline q}J_{\underline m \underline
p}+\eta_{\underline m \underline p}J_{\underline n \underline
q}
\eqno(2.1)$$

$$[ P_{\underline n },J_{\underline p \underline q}]
=+\eta_{\underline n \underline p}P_{\underline q}
-\eta_{\underline n \underline q}P_{\underline p}
\eqno(2.3)$$
\par
We can write the coset representatives in the form
$$g(X,\phi)= exp(X^{n}P_{n}+X ^{n'}P_{n'})
exp(\phi^{n m'}J_{n m'}+\phi ^{n'm'}J_{n'm'})
$$
$$\equiv
exp(X^{\underline n}P_{\underline n})
exp(\phi\cdot J)
\eqno(2.4)$$
We distinguish $X$ from $X'$ by the indices they carry,
in other words the prime on the $X$ is understood to be present and we
just write $X ^{n'}$.
The Cartan forms are given by
$$g^{-1}dg\equiv e^n P_n +f^{n'}P_{n'}+ \Omega^{n m'}J_{n m'}
+\Omega^{n'm'}J_{n'm'}+w^{nm}J_{nm}
\eqno(2.5)$$
which we may express as
$$g^{-1}dg=
exp(-\phi\cdot J) (dX^{\underline n} P_{\underline n}) exp(\phi\cdot J)
+exp(-\phi\cdot J)d exp(\phi\cdot J)
\eqno(2.6)$$
A straightforward calculation shows  that
$$e^n=d X^{\underline m}\Phi_{\underline m}{}^{n}=
-2\phi^{nm'}d X_{m'}+dX^{n}+\dots
$$
$$
f^{n'}=dX^{\underline m}\Phi_{\underline m}{}^{n'}=
 2\phi^{m'n'}d X_{m'}-2\phi^{n'm}dX_m+d X^{n'}
$$
$$
\Omega ^{nm'}=d\phi^{nm'}+2   (-\phi^{p'm'}d\phi_{p'}^{\ n}
+\phi^{p' n}d\phi_{p'}^{\ m'})$$
$$
\Omega ^{n'm'}=d \phi^{n'm'}+((\phi^{p'n'}d\phi_{p'}^{\ m'}+
\phi^{ p n'}d\phi_{p}^{\ m'} -(n' \leftrightarrow m')),\ \
w^{nm}=
(\phi^{p' n}d\phi_{p'}^{\ m}-(n \leftrightarrow m))
\eqno(2.7)$$
where $\Phi_{\underline n}{}^{\underline m}$ is defined by
$exp(-\phi\cdot J) P_{\underline n} exp(\phi\cdot J)=
\Phi_{\underline n}{}^{\underline m}P_{\underline m}$ and $+\ldots $
means higher order terms in
$\phi^{\underline n\underline m}$.
\par
Under a group transformation, $g_0 g(X,\phi)= g (\hat X, 
\hat \phi ) h_0$, the Cartan forms transform by the
$J_{nm}$ transformations with associated parameter $r^{nm}$  in accord
with equation (1.4). The
effect of the
$P_{\underline n}$ transformations in
$g_0$ is to simply transform the $X^{\underline n}$. Writing
$h_0= 1+r^{nm} J_{nm}$. we find that
$$
\hat e^n =e^n-e^p r_p^{\ n},\ \ \hat f^{n'}=f^{n'},
\hat \Omega^{nm'}=  \Omega^{nm'}+ 2 r^{p n}\Omega_{p}^{\ \ m'}-
 2 r^{p m'}\Omega_{p}^{\ \ n}\eqno(2.8)$$
$$
\hat w^{nm}=  w^{nm}-2(r^{np} w^{\ m}_p -
(n\leftrightarrow m) ) +dr^{nm}
\eqno(2.9)$$
to lowest order in $r^{nm}$.
\par
Clearly, we can set $f^{n'}=0$ and
preserve SO(1,D-1) and reparameterisation symmetry. At the linearized
level, examining equation (2.7)  we find   it implies that
$d X ^{n'}=2\phi^{n'}_{\ m}dX^m$ or
$$ 2\phi^{n'}_{\ m}={\partial \xi^p\over \partial X^m}
{\partial X^{n'}\over \partial \xi^p}
\eqno(2.10)$$
If we choose static gauge this equation becomes,
$$2\phi^{n'}_{\ m}={\partial
X^{n'}\over \partial X^m},
\eqno(2.10)$$
While solving for $f^{n'}=0$
to   all orders may be complicated  it is clear that its
content is to solve for $\phi^{n'}_{\ m}$ in terms of $\partial_m X
^{n'}$.
\par
What is really of interest to us is the non-linear form of
$e^n$ once we have solved this constraint. Examining equation (2.7) we
find that  the vector $f^{\underline n}\equiv (e^n, f^{n'})$
is related by a Lorentz
transformation
 to the vector $(dX^n, dX^{n'})= d X^{\underline n}$. As such,
$$e_p^{\ n}\eta _{n m}e_q^{\ m}=\partial _p X^{\underline n}
\partial _q X^{\underline m} \eta _{\underline n \underline m}
\eqno(2.11)$$
since $f^{n'}=0$. The above expression is invariant under
$g_0$ transformations. A worldvolume reparameterisation
and group invariant action is therefore given by
$$ \int d^p\xi \ det e= \int d^p\xi \sqrt {-det (\partial _p
X^{\underline n}
\partial _q X^{\underline m} \eta _{\underline n \underline m})}
\eqno(2.12)$$
In other words,  the well known generalisation of the Nambu action for
the string to a p-brane follows in a straightforward consequence of
taking the non-linear realization $ISO(D-1,1)$ with subgroup $SO(p,1)$.
\par
Clearly, had we not included the
Lorentz group in our coset and
introduced the corresponding Goldstone bosons the veilbein on the
brane would have been trivial and we would not have found the
above action.
\medskip
{\bf{3. Superbranes}}
\medskip
For superbranes we take the supergroup $\underline G$ to be that 
 considered in [15] and applied there to the context of M theory.
In accord with section
one it has the two subgroups $\underline K$ and $\underline H$.
The generators of
$\underline K$ obey the relations
$$\{Q_{\underline \alpha} ,Q_{\underline \beta} \}
=Z_{\underline \alpha\underline \beta},\qquad
[Q_{\underline \gamma},Z_{\underline \alpha\underline \beta}] = 0,
\qquad
[Z_{\underline \alpha\underline \beta},
Z_{\underline \gamma\underline \delta}]=0,
\eqno(3.1)$$
while those of  the
automorphism group
$\underline H$ satisfy
$$
[Q_{\underline \alpha}, R_{\underline \gamma}{}^{\underline \delta}]=
 \delta_{\underline \alpha}{}^{\underline \delta} Q_{\underline
\gamma},\  [Z_{\underline \alpha \underline \beta}, R_{\underline
\gamma}{}^{\underline \delta}]
= \delta_{\underline \alpha}{}^{\underline \delta}
Z_{\underline \gamma
\underline \beta} + \delta_{\underline \beta}{}^{\underline \delta}
Z_{\underline \alpha \underline \gamma}.
\eqno(3.2)$$
The generators $Z_{\underline \alpha\underline \beta}$ are Grassmann
even generators labeled by spinor indices.  They  are obviously
symmetric in these indices.  It is easy to verify that
such an algebra obeys the
generalized super Jacobi identities.  Let us
assume that the generators
$Z_{\underline
\alpha\underline \beta}$ form the most general symmetric matrix.
Expanding this matrix out in terms of the enveloping algebra of the
generators of the
relevant Clifford algebra we
find that $Z_{\underline \alpha\underline \beta}$ contains a set of
totally anti-symmetric tensorial generators which constitute the
central charges:
$$
Z_{\underline \gamma}{}^{\underline \delta}= \sum_p \sum_{\underline
n_1\ldots
\underline n_p} (\gamma^{\underline n_1\ldots
\underline n_p}C^{-1})_{\underline \gamma}{}^{\underline \delta }
Z_{\underline n_1\ldots \underline n_p}.
\eqno(3.3)$$
The central charges include the momenta $P_{\underline n}\equiv
Z_{\underline n}$. Both the eleven dimensional and the six dimensional
superalgebras associated with the fivebrane equations of motion
have a supersymmetry algebra in which the central charges form
the most general matrix [18,15]. In the latter case one must take
an appropriate index range for the indices
$\underline \alpha , \underline \beta \ldots$.
 The corresponding
superalgebras in the IIA and IIB theories for which the
$Z_{\underline \alpha\underline \beta}$ form the most general matrix in ten
dimensions are well
known.
\par
The automorphism group is $GL(c_d)$ where $c_d$ is
the number  of supercharges. The precise properties of the matrices
under complex conjugation being given by implementing the Majorana
or other properties of the spinorial supercharge on the commutator
relation of the supercharge with the automorphisms.  For the case of
the eleven dimensional superalgebra it is $GL(32)$
\par
To gain a more familiar set of generators
we may expand $R_{\underline \gamma}{}^{\underline \delta}$ out in
terms of the elements of the enveloping Clifford algebra
$$
R_{\underline \gamma}{}^{\underline \delta}= \sum_p \sum_{\underline
n_1\ldots
\underline n_p} (\gamma^{\underline n_1\ldots
\underline n_p})_{\underline \gamma}{}^{\underline \delta }
R_{\underline n_1\ldots
\underline n_p}.
\eqno(3.4)$$
\par
If   the only central charge is the momentum then
the right-hand side of the anti-commutator of two
supercharges is
$\gamma^{\underline n}P_{\underline n}$. In this case,  the most
general automorphism group is  by definition the spin group.
Hence, the automorphism group $GL(32)$
is the natural generalisation to  branes
of the spin group relavent to point particles.    In
the above expansion the  generators
$R_{\underline n \underline m}$ are those of the Spin Lie algebra.
\par
The automorphism group has been found to be relavent to
the   dynamics of the fivebrane, In particular,  its covariant
equations of motion [4]  have been found [15] to possess additional
conserved currents beyond those  associated with Spin
transformations, central charges and supersymmetry. These new
currents lead to conserved charges that generate, at the classical linearized
level, a  subalgebra of the automorphism group.
\par
We may also consider a subgroup of the full automorphism group.
One such   is given by the set of symmetric
matrices, these form the Lie algebra of  Sp($c_d$).
The generators of this
subalgebra are $S_{\underline \gamma\underline \delta}=
R_{\underline \gamma\underline \delta}+R_{\underline \delta\underline
\gamma}$ where
$R_{\underline \gamma\underline \delta}=
R_{\underline \gamma}{}^{\underline \beta}(C^{-1})_{\underline \beta
\underline \delta}$.  Clearly, in the decomposition of equation
(3.4)  this means keeping only those terms for which the
symmetric matrices enter.  In eleven dimensions these are the
generators
$R_{\underline n_1\ldots \underline n_p}$ of ranks one, two and five.
The commutator of the
generators $S_{\underline \gamma\underline \delta}$ with
the supercharges is given by
$$
[Q_{\underline \alpha}, S_{\underline \gamma \underline \delta}]=
(C^{-1})_{\underline \alpha \underline \delta}
 Q_{\underline \gamma}+ (C^{-1})_{\underline \alpha \underline \gamma}
 Q_{\underline \delta}.
\eqno(3.5)$$
\par
We divide the  generators in $\underline K$ into
$Q_{\underline \alpha}= (Q_\alpha ,Q_{\alpha'})$
and
$Z_{\underline \alpha \underline \beta}=( Z_{ \alpha \beta}, Z_{
\alpha \beta'}, Z_{ \alpha' \beta'})$ and the generators of
$\underline H$ as $R_{\underline \gamma}{}^{\underline \beta}
=(R_{ \gamma}{}^{ \beta},R_{ \gamma'}{}^{ \beta},
R_{ \gamma}{}^{ \beta'},R_{ \gamma'}{}^{ \beta'})$.
We take the Lie superalgebra of the group $G$ to contain the
generators
$Q_\alpha$ and $ Z_{\alpha \beta}$
 as well as some  of the generators
$  R_{\gamma}{}^{\delta}$
\par
Initially,  we will take the non-linear realization approach to the
superbrane in which  the coset representatives
 are parameterised by the coordinates on the brane and the
transverse fields. Proceeding in  this way, we will find the
expressions for the superbrane in static gauge.
  As such, the coset
representatives of
$\underline G/H$ are  written as
$$g= e^{X^{\alpha \beta}Z_{\alpha \beta}+\theta ^\alpha Q_\alpha}
e^{X^{\alpha \beta'}Z_{\alpha \beta'}+X^{\alpha' \beta'}Z_{\alpha'
\beta'}+\Theta ^{\alpha'} Q_{\alpha'}}e^{\phi '\cdot R '}
\eqno(3.6)$$
where $\phi'\cdot R'$ is a sum that includes all the generators in
$\underline H$, except for those in $H$. The Cartan forms are given by
$$g^{-1}d g= e^{-\phi'\cdot R'}d
e^{\phi'\cdot R'}+
$$
$$e^{-\phi'\cdot
R'}(Y^{\alpha\beta}Z_{\alpha\beta}
+Y^{\alpha\beta'}Z_{\alpha\beta'}
+Y^{\alpha'\beta'}Z_{\alpha'\beta'}+
d\theta ^{\alpha} Q_{\alpha }
+d\Theta ^{\alpha'} Q_{\alpha' })
e^{\phi'\cdot R'}
$$
$$\equiv  F^{\alpha\beta}Z_{\alpha\beta}
+F^{\alpha\beta'}Z_{\alpha\beta'}
+F^{\alpha'\beta'}Z_{\alpha'\beta'}+
F^{\alpha} Q_{\alpha } + F ^{\alpha'} Q_{\alpha' }
+e^{-\phi'\cdot R'}d
e^{\phi'\cdot R'},
\eqno(3.7)$$
where
$Y^{\alpha\beta}=dX^{\alpha\beta}+{1\over
4}(\theta^{\alpha}d\theta^{\beta}+\theta^{\beta}d\theta^{\alpha})$,
$Y^{\alpha\beta'}=
dX^{\alpha\beta'}-d\theta^{\alpha}d\Theta^{\beta'} $ and
$Y^{\alpha'\beta'}=dX^{\alpha'\beta'}+{1\over
4}(\Theta^{\alpha'}d\Theta^{\beta'}+\Theta^{\beta'}d\Theta^{\alpha'})$
\par
In order to gain some understanding of  the physical content of the
formalism we will begin by giving a linearized analysis. To lowest
order in
$\phi_{\underline
\delta }{}^{\underline
\gamma}$ the Cartan forms are given by
$$
F^{\alpha}= d\theta ^{\alpha }+ d\Theta ^{\delta'}
\phi_{\delta '}{}^{\alpha}+
d\theta ^{\delta}
\phi_{\delta }{}^{\alpha},\
F^{\alpha \beta }= Y^{\alpha \beta }+
Y^{\alpha \delta' }\phi_{\delta '}{}^{\beta}
+Y^{\alpha \delta }\phi_{\delta }{}^{\beta}
+Y^{\beta \delta }\phi_{\delta }{}^{\alpha},
$$
$$
F^{\alpha '}= d\Theta ^{\alpha '}+ d\Theta ^{\delta'}
\phi_{\delta '}{}^{\alpha '} +
d\theta ^{\delta}
\phi_{\delta }{}^{\alpha '},$$
$$
F^{\alpha '\beta '}= Y^{\alpha '\beta '}+
Y^{\alpha '\delta' }\phi_{\delta '}{}^{\beta '}+
+Y^{\beta '\delta '}\phi_{\delta '}{}^{\alpha '}
+Y^{\alpha '\delta
}\phi_{\delta }{}^{\beta '} +Y^{\beta '\delta }\phi_{\delta
}{}^{\alpha '},\
$$
$$
F^{\alpha \beta '}= Y^{\alpha \beta ' }+
Y^{\alpha \delta' }\phi_{\delta '}{}^{\beta '}+
+Y^{\beta '}\phi_{\delta '}{}^{\alpha }
+Y^{\alpha \delta  }\phi_{\delta }{}^{\beta '} +Y^{\delta \beta '
}\phi_{\delta }{}^{\alpha }
\eqno(3.8)$$
\par
We wish to place constraints on  the Cartan forms
that respect the group $\underline G$ and lead to the equations
of motion of the superbrane.
Taking the Goldstone fields to be of   order
one and the coordinates $X^{\alpha \beta}$ and $\theta^{\alpha}$
to be of order zero, we will first carry out
a linearized analysis.  As
explained in section one,  the equation of motion are given by setting
certain of the covariant
derivatives in  the Goldstone fields to vanish. As such, we will only require
the supervielbein to order zero. It is straightforward to read it off
from equation (3.8) and we find that
$$
F= \left(\matrix{{1\over 2}(\delta ^{\alpha
}_{\gamma}\delta_{\delta}^{\beta}
+\delta ^{\beta
}_{\gamma}\delta_{\delta}^{\alpha})
&0\cr -{1\over 4}(\delta ^{\alpha
}_{\gamma}\theta^{\beta}+
\delta ^{\beta }_{\gamma}\theta^{\alpha})&\delta ^{\alpha
}_{\gamma}\cr}\right)
\eqno(3.9)$$
\par
We can also read off the linearized Cartan forms from equation (3.8)
and,  multiplying the result by the inverse supervielbein,  we find that
the covariant derivatives of the Goldstone fields are given by
$$\nabla_{\alpha}\Theta^{\alpha'}=D_{\alpha}\Theta^{\alpha'}+
\phi_{\alpha}{}^{\alpha'},\
\nabla_{\gamma\delta}\Theta^{\alpha'}
=\partial_{\gamma\delta}\Theta^{\alpha'},$$
$$\nabla_{\gamma}X^{\alpha\beta '}= D_{\gamma}X^{\alpha\beta '}
-\delta_{\gamma}^{\alpha}\Theta^{\beta '},\
\nabla_{\gamma\delta} X^{\alpha\beta '}= \partial_{\gamma\delta}
X^{\alpha\beta '} + {1\over 2}(\delta_{\gamma}^{\alpha}
\phi_{\delta}{}^{\beta'} + \delta_{\delta}^{\alpha}
\phi_{\gamma}{}^{\beta'})
\eqno(3.10)$$
where
$$D_{\alpha}\equiv\partial _{\alpha}+{1\over
2}\theta^{\delta}\partial_{\gamma\delta}, \
\partial_{\gamma\delta}\equiv {\partial \over
\partial {X^{\gamma\delta}}}
\eqno(3.11)$$
It is a consequence of the linearized analysis that
$\phi_{\gamma}{}^{\beta'}$  is the only part of
$\phi_{\underline \gamma}{}^{\underline \beta}$ that enters any of
these expressions. The covariant derivatives of the other Goldstone
fields will not interest us here.
\par
We now choose the group $H$ to be just the Lorentz group. The
 Cartan forms carry a very reducible
representation with respect
to this subgroup and hence with respect to $\underline
G$. Consequently, we may choose many of the covariant derivatives of the
Goldstone fields to vanish. In fact,  we will choose the set
$$\nabla_{\gamma}\Theta^{\alpha'}=0,
\eqno(3.12)$$
$$ \nabla_{\gamma}X^{\alpha\beta '}(\gamma^{n'}C^{-1})_{\alpha\beta '}
=0
\eqno(3.13)$$
and
$$ \nabla_{\gamma\delta}\Theta^{\alpha'}-{2\over r_p}
(\gamma^{n}C^{-1})_{\gamma\delta}(\gamma^{n}C^{-1})^{\alpha\beta }
\nabla_{\alpha\beta }\Theta^{\alpha'}=0
\eqno(3.14)$$
where $r_p$ is the number of supercharges in $G$. The
superspaces defined by the above coset space construction
contain many central charge associated coordinates in addition to
the usual Minkowski coordinates. The Grassmann even coordinates of the
background
superspace $\underline G/ \underline H$ include the
coordinates $X^{\underline n}\equiv  X^{\underline \alpha\underline \beta
'}(\gamma^{ \underline n}C^{-1})_{\underline\alpha\underline\beta
}$ of Minkowski space, but  also all the other
coordinates corresponding to the
expansion of the supercharge of equation (3.3). Similarly,
the coset space $G/H$ on which
the above superfields are defined includes  the Minkoswki space
coordinates $X^n\equiv  X^{\alpha\beta '}(\gamma^{n}C^{-1})_{\alpha\beta }$
as well as a  corresponding number of other Grassmann even
coordinates associated with the central charges in $G$ other than the momentum.

Equation (3.14) sets the central
charge dependence of
$\Theta ^{\beta'}$ to be trivial except for the dependence on the
coordinates $X^n$. Equation (3.13) then implies  that $X^{n'}$ has the
same  dependence.  As such, the above equations are equivalent to  the
conditions
$$D_\alpha \Theta ^{\beta'}= -\phi_{\alpha}{}^{\beta'}
\eqno(3.15)$$
and
$$ D_\gamma X^{n'}=(\gamma^{n'}C^{-1})_{\gamma\delta' }\Theta^{\delta'}
\eqno(3.16)$$
where all the superfields $\Theta ^{\beta'}$,  $X^{n'}$ and
$\phi_{\alpha}{}^{\beta'}$ depend only
on
$X^n$ and $\theta^{\alpha}$.
It is known [17] from the superembedding approach that these
equation imply the correct field content and equations of motion for
most of the  superbranes including the  2 and 5-branes of M theory.
\par
Let us study the consequences of these equations for the
branes in M theory. We must decompose the eleven dimensional Clifford
algebra to one that keeps manifest the Clifford algebra appropriate
to the brane. This results in a corresponding decomposition of the
spinor index $\underline \alpha= (\alpha, i)$. For the fivebrane
$\alpha=1,\dots ,4$ are the Weyl projected spinor indices of
Spin(1,5) and $i=1,\dots ,4$ are the indices of the internal group
$Usp(4)=Spin(5)$ while for the twobrane
 $\alpha=1,2$ are the  spinor indices of
Spin(1,2) and $i=1,\dots ,8$ are indices of $Spin(8)$.
The analysis of equation (3.16) depends only on the properties of
this decomposition of the eleven dimensional Clifford algebra. For the
fivebrane of M theory it implies that [17]
$$ D_{\alpha i}\Theta _\beta {}^{ j}\sim
-(\rlap/ \partial)_{\alpha \beta}(\gamma_{n'})_i{}^{j} X^{n'}
+\delta ^j_i (\gamma_{n_1n_2n_3})_{\alpha \beta} h^{n_1n_2n_3}
\eqno(3.17)$$
where $h_{n_1n_2n_3}$ is the self-dual gauge field strength of the
fivebrane. At the linearized level,
the field $h_{n_1n_2n_3}$  obeys the Bianchi identity implying it is the
curl of a second rank gauge field in addition to
being self-dual. This relation   follows from the constraint of
equation (3.16) by applying further covariant derivatives.
\par
For the twobrane we find that
$$ D_{\alpha i}\Theta _{\beta}{}^{ j}\sim
-(\rlap/ \partial)_{\alpha \beta}(\gamma_{n'})_i{}^{j} X^{n'}
\eqno(3.18)$$
\par
Equation (3.15) implies that  the Goldstone field
$\phi_{\alpha}^{\beta'}$ associated with the automorphism group
$\underline H$ is non-vanishing. Expanding
$$
\phi_{\underline \gamma}{}^{\underline \delta}=
\sum_p \sum_{\underline
n_1\ldots
\underline n_p} (\gamma^{\underline n_1\ldots
\underline n_p})_{\underline \gamma}{}^{\underline \delta }
\phi_{\underline n_1\ldots
\underline n_p}.
\eqno(3.19)$$
we find that both for the fivebrane and the twobrane that
$\phi_{n}{}^{m'}\sim\partial_n X^{m'}$ which is the same as  the
condition of equation (2.10) for the bosonic brane. It relates
the Goldstone fields for the broken Lorentz generators
to the derivative of Goldstone fields of broken translations.
However, for the fivebrane we find in addition that the field
strength for the self-dual gauge field is related to the
Goldstone fields for broken generators of the automorphism group
GL(32), in particular $h_{n_1n_2n_3}\sim \phi_{n_1n_2n_3}$. We note
that these generators are the
$R_{nmp}$ contained in equation (3.4) and that these generators are not
contained in the  subgroup $Sp(32)$. This is consistent with the
existence of the additional rank three conserved current found [15] in the
covariant equations of motion of the fivebrane.
\par
The non-linear realization described above contains more Goldstone
superfields
than those of equations (3.15) and (3.16). One could consider imposing
more constraints generalizing that of equation (3.16) to include the
covariant derivatives of the Goldstone fields associated with the
central charges. These would determine  these additional fields
in terms of the fields that we have already. In particular,  one could
impose a constraint the implies that the Goldstone field associated
with the central charge $Z_{nm}$ is the second rank gauge field of the
fivebrane.
\par
It is not clear that the choice of subgroup $H$ and the related
constraints of equations (3.12) to (3.14)
are the unique choice that leads to the correct
dynamics. Indeed, it would be interesting to explore other choices.
In particular,
one could also consider constraints which allow the superfields to
depend on the central charges of the six dimensional subalgebra, but
in a rather restricted way.
\par
To make contact with the superembedding formalism,  and to carry out an
all orders analysis, it will be
convenient to adopt a different parameterisation
of the coset representatives. As explained in section one,  we can take the
Goldstone fields to depend on the coordinates of the superembedded
manifold and treat the transverse and longitudinal coordinates of
the superbrane on an equal footing.  This allows us to work in a
superreparameterisation invariant manner and not choose superstatic gauge
as above.
As such, we choose our coset representatives to be
$$g= e^{X^{\underline \alpha \underline \beta}
Z_{\underline\alpha \underline\beta}+\Theta
^{\underline \alpha} Q_{\underline \alpha}} e^{\phi'\cdot R'}
\eqno(3.20)$$
Where the coordinates $Z^{\underline M}
\equiv (X^{\underline \alpha \underline \beta}, \Theta ^{\underline
\alpha})$, are not be confused with the central charges
$Z_{\underline\alpha \underline\beta}$,
 depend on the coordinates,
$z^{M}=(\xi^{\alpha\beta}, \theta^{\alpha})$ that parameterise the
superbrane. The transformation
from the fields of   equation (3.6) to that in the above equation is
obtained by choosing superstatic gauge $X^{\alpha
\beta}=\xi^{\alpha\beta}, \
\Theta ^{\alpha}=\theta ^{\alpha}$
 and making the shift
$$X^{\alpha \beta'}\to
X^{\alpha\beta'}+{1\over 2}\theta^\alpha \Theta ^{\beta'},
\eqno(3.21)$$
with all the other fields being the same.
Constructing the Cartan forms as before,  we find that
$$g^{-1}d g=
 e^{-\phi'\cdot
R}( E^{\underline A} K_{\underline A})
e^{\phi'\cdot R'}+
e^{-\phi'\cdot R'}d
e^{\phi'\cdot R'}$$
$$\equiv F^{\underline A} K_{\underline A}+
e^{-\phi'\cdot R'}d e^{\phi'\cdot R'}
\eqno(3.22)$$
In this equation the index $\underline A
=(\underline \gamma\underline \delta, \underline \gamma)$ and the
one forms $E^{\underline A}\equiv dZ^{\underline \Pi}
E_{\underline M}{}^{\underline A}$ contain  the supervielbeins
$E_{\underline M}{}^{\underline A}$ of the
background superspace and have the form
$$
E_{\underline M}{}^{\underline A}
= \left(\matrix{{1\over 2}( \delta ^{\underline \alpha
}_{\underline\gamma}\delta_{\underline \delta}^{\underline\beta}
+\delta ^{\underline\beta
}_{\underline\gamma}\delta_{\underline\delta}^{\underline\alpha})
&0\cr -{1\over 4}(\delta ^{\underline\alpha
}_{\underline\gamma}\Theta^{\underline\beta}+
\delta ^{\underline\beta
}_{\underline\gamma}\Theta^{\underline\alpha})&\delta
^{\underline\alpha }_{\underline\gamma}\cr}\right)
\eqno(3.23)$$
Defining $e^{-\phi'\cdot R'}( K_{\underline A}) e^{\phi'\cdot R'}
\equiv \Phi_{\underline A}{}^{\underline B} K_{\underline
B}$ we find that
$$
F^{\underline A}
= E^{\underline B}\Phi_{\underline B}{}^{\underline A}
\eqno(3.24)$$
\par
The dynamics of the non-linear theory should be given by imposing the
constraints of equations  (3.12) to (3.14)  where the derivatives are
now
defined for the full  theory. In the next section we will argue
that these constraints  correctly encode  the non-linear dynamics of
the brane. It would be interesting to examine the implications of these
constraints in detail and to see in particular what they imply for all
the
Goldstone fields associated with the automorphism. For those
superbranes that have a straightforward action, it is possible that
the supervielbein on the superbrane that arises in the non-linear
realization will play an essential role in defining the action as it
does for the case of  the bosonic brane considered in section one.
\medskip
{\bf {4. Relation to the Superembedding Formalism}}
\medskip
In the superembedding approach [7] we consider one supermanifold
$M$ embedded in another $\underline M$. The preferred vector fields on
the former are given by $E_{A}=E_{A}{}^{M}\partial _{M}$,  where
$E_{A}{}^{M}$ are the inverse supervielbein, while those on the latter
are $E_{\underline A}=E_{\underline A}{}^{\underline M}
\partial _{\underline M}$. The fields on $M$ must point somewhere in
$\underline M$ as is expressed by the equation
$$E_{A}{}^{M}\partial _{M}=E_{A}{}^{\underline A}
E_{\underline A}{}^{\underline M}
\partial _{\underline M}.
\eqno(4.1)$$
The assumption in the superembedding  approach is that
$$E_{\alpha}{}^{\underline a}=0.
\eqno(4.2)$$
This states that  the Grassmann odd preferred
vector fields in $M$ can be expressed in terms of only the
Grassmann odd vector fields in $\underline M$. Although the elegant
simplicity of this assumption is apparent, its deeper meaning is still
unclear.
For many branes, and certainly many
of the most interesting ones, this single assumption is sufficient to
put the theory on-shell and  determine its complete equations
of motion [17]. This is the case for the twobrane and fivebrane of M
theory.
\par
We may re-express equation (4.1) as
$$dz^{M}E_{M}{}^{A} E_{A}{}^{\underline B}=
dz^{M}\partial _{M}
Z^{\underline \Lambda} E_{\underline \Lambda}{}^{\underline B}
\eqno(4.3)$$
Multiplying by $\Phi_{\underline
B}{}^{\underline C}$, identifying the
supervielbien on the embedded super manifold with the Cartan form in
the $K_A$ directions, i.e.  $E_M{}^{A}=F_M{}^{A}$,  and
recognizing $F^{\underline A}$ from equation (3.22) we conclude that
$$F_{A}{}^{\underline B}\equiv E _{A}{}^{M}
F_{M}{}^{{\underline B}}= \nabla_A Z^{\underline B}=
E_A{}^{\underline C}\Phi_{\underline C}{}^{\underline B}
\eqno(4.4)$$
The $\Phi$ defined above equation (3.24) are constructed from the Goldstone
fields associated with the automorphism group. They  form an
invertible matrix and also must have indices such that they are always
Grassmann even; implying for example that $\Phi_{\underline
\beta}{}^{\underline b}=0$.
It follows from equation (4.4) that  the inverse Higgs condition
$$F_{\alpha}^{\underline a}=0,
\eqno(4.5)$$
is the  necessary and sufficient
condition for the embedding condition of equation (4.2). However, the
superembedding formalism uses superfields that do not depend on the
central charges other than the usual coordinates of Minkowski space. 
Hence, provided we adopt the inverse Higgs conditions of equations
(3.12),
(3.13) and  (3.15) we should find that they describe the dynamics of
the full non-linear theory. This follows because
 equation (3.13) is just the condition of
equation (4..5), equation (3.14) eliminates the dependence of the
superfields on the coordinates associated with the central charges and
equation (3.12) identifies the automorphism Goldstone fields in terms
of the fields of the brane. We note that the first two 
inverse Higgs conditions can be written as $F_{\alpha}{}^{\underline A}
=\delta _\alpha^{\underline A}$.
\par
The relationship between the inverse Higgs mechanism and the
superembedding condition has been previously discussed in the context
of superbranes in lower dimensions and their construction as
non-linear realizations [14], [20]. However,  in these cases  the
dynamics of the branes are not determined by the embedding condition
of equation (4.5). It would be interesting to apply the methods
developed in this paper to  branes of this type.
\par
Identifying  the supervielbien $F_{A}^{M}$ in the
non-linear realization approach with the supervielbein of the
superembedding approach implies that $F_{A}{}^{B}=\delta_{A}{}^{B}$
and, as a consequence, 
 $E_{A}{}^{\underline B}\Phi_{\underline B}{}^{B}=\delta_{A}{}^{B}$.
Taking fermonic indices,  and including the above condition of equation
(4.5),  we conclude that
$E_{\alpha}{}^{\underline \beta}\Phi_{\underline
\beta}{}^{\gamma}=\delta_{\alpha}{}^{\gamma}$. We can however choose
a set of frames normal to the superbrane. By expressing these
vector fields in terms of the preferred frames of the background
superspace, we can define an extension of the $E_{A}{}^{\underline B}$
to $E_{\underline A}{}^{\underline B}$. Making this extension in an
appropriate way we conclude that $E_{\underline A}{}^{\underline
B}$ is the inverse of the
$\Phi_{\underline A}{}^{\underline B}$. Thus the Goldstone fields
associated with the automorphism group specify  the
relation between preferred frames associated with the branes and those
in the background superspace.
\par
Although the superfields do not in the end depend on the
central charges,  without the central charges in the original algebra
one would not have the large automorphism group which is essential
for the theory to contain the fields of the brane. This is apparent
as a consequence of  the discussion of the above paragraph, however,
another way to observe this fact is to note that the supervierbeins on
the embedded manifold would be essentially trivial if we had not
included the Goldstone fields for the automorphism group.
\medskip
{\bf {5. Conclusion}}
\medskip
We have  shown that the branes of M theory and, by reduction, those in
ten dimensions can be described by the theory of non-linear
realization provided we use a superalgebra that has the large
automorphism group considered in [15]. Indeed,  the field strengths of
the worldvolume fields of the  brane arise as  the Goldstone bosons
for  the automorphisms.   It is a widely held
belief  that branes correspond to
the low energy effects  of a defect in space-time the breaks
certain symmetries. We have shown in this paper that their
dynamics is essentially determined by a knowledge of which symmetries
are broken.
\par
In reference [19] it was explained how  the fields of the branes
arise as zero modes of the the corresponding background supergravity
theory.
It would  be interesting understand the connections between that work
and that of this paper.  It would also be interesting to find if
there are connections between the non-linear realization proposed
in this paper and the coset consider in  reference [14] which was
based  on the Osp groups.
\par
The necessity of including at least parts of the GL(32) automorphism
group in the construction of the non-linear realization of the branes
of M theory supports the idea proposed in [15] that this symmetry may
play an important role in M theory.
\medskip
{\bf {Acknowledgment}}
\medskip
The author would like to thank Jon Bagger for
a number of  useful discussions at an early stage of this work and
the hospitality of the Ecole Normale Superieure, where some of this
work  was carried out.  This work was
supported in part by the EU network  on Integrability,
Non-perturbative effects, and Symmetry in Quantum Field theory
(FMRX-CT96-0012).

\medskip
{\bf {References}}
\medskip
\parskip 0pt
\item{1}
W.Siegel, { Phys. Lett. \/} {\bf B128} (1983) 397.
\item {2}
E. Bergshoeff, E. Sezgin and P. K. Townsend,
{  Phys. Lett.\/} {\bf 189B} (1987) 75; Ann. of Phys. 185 1988 330.
\item{3}
M. Cederwall, A. von Gussich, B. E. W. Nilsson, A. Westerberg,
 Nucl.Phys.
{\bf  B490} 1997 163;\
M. Cederwall, A. von Gussich, B. E. W. Nilsson, P. Sundell and
A.  Westerberg,\
{ Nucl.Phys.} {\bf B490} (1997) 179;\
M. Aganagic, C. Popescu, J.H. Schwarz, Phys. Lett. {\bf B393} 1997
311; E. Bergshoeff, P.K. Townsend,
{ Nucl.Phys.} {\bf B490} (1997) 145;
M. Aganagic, J. Park, C. Popescu and J. H. Schwarz,
{\sl Nucl.  Phys.} {\bf B496} (1997) 191.
\item{4}
P. S. Howe, E. Sezgin, and P. C. West.
 Covariant field equations of the {M} theory five-brane.
 { Phys. Lett.}, B399:49--59, 1997. hep-th/9702008.
\item{5}
I. Bandos, K. Lechner, A. Nurmagambetov, P.
Pasti, D. Sorokin and M.  Tonin, {\sl Phys. Rev. Lett.} {\bf 78} (1997)
4332;
I.  Bandos, K. Lechner, A.
Nurmagambetov, P.  Pasti, D. Sorokin and M. Tonin,
{ Phys. Lett.\/} {\bf B408} (1997) 135.
\item{6}
M. Perry and J. H. Schwarz, { Nucl. Phys.\/} {\bf 498} (1997) 47;\
J. H. Schwarz, { Phys. Lett.\/} {\bf B395} (1997) 191;
Mina Aganagic, Jaemo Park, Costin Popescu, John H. Schwarz,
World-Volume Action of the M Theory Five-Brane, hep-th/9701166,
Nucl.Phys. B496 (1997) 191-214
\item{7} For two  reviews  see
P. S. Howe, E. Sezgin and P. C. West,
{ Aspects of Superembeddings\/},
D. V. Volkov Memorial Volume,
{ Lecture notes in physics \/}, Vol. 509, p. 64,
Springer--Verlag, Berlin, Heidelberg 1998.
(1988) 709; and
D. Sorotkin, { Superbranes and  Superembeddings\/},
hep-th/9906142.
\item{8} J. Hughes, J. Polchinski, Nucl. Phys. {\bf B 278} (1986)
147;  J. Hughes, J. Liu, J. Polchinski, Phys. Lett. {\bf B 180}
(1986) 370.
\item{9} S. Coleman, J. Wess and B. Zumino, { Phys. Rev.\/}
{\bf 177} (1969) 2239;\
K. Callan, S. Coleman, J. Wess and B. Zumino,
{ Phys. Rev.\/} {\bf 177} (1969) 2247.
\item{10}
D. V. Volkov, { Sov. J. Part. Nucl.\/} {\bf 4} (1973) 3;
D. V. Volkov and V. P. Akulov, { JETP Letters} {\bf 16} (1972) 438;
{ Phys. Lett. \/} {\bf B46} (1973) 109.
\item{11} J. Bagger, A. Galperin, Phys. Lett. {\bf B 336}
(1994) 25;  Phys. Rev. {\bf D 55} (1997)
1091;  Phys. Lett. {\bf B 412} (1997)
296.
\item{12}
F. Gonzalez-Rey, I.Y. Park, M. Ro\v{c}ek, Nucl. Phys. {\bf B 544}
(1999) 243;
 M. Ro\v{c}ek, A. Tseytlin, Phys. Rev. {\bf D 59} (1999)
106001. S. Bellucci, E. Ivanov, S. Krivonos, Phys. Lett. {\bf B
460} (1999) 348; E. Ivanov, S. Krivonos, Phys. Lett. {\bf B 453}
(1999) 237.
\item{13}
S. Belucci, E. Ivanov and S. Krivonos, { Partial
breaking $N=4$ to $N=2$: hypermultiplet as a Goldstone superfield\/},
hep--th/9809190.\
S. Belucci, E. Ivanov and S. Krivonos, { Partial breaking of
$N=1$, $D=10$ supersymmetry}, hep--th/9811244;\
F. Gonzalez--Rey, I. Y. Park and M. Ro\~cek, { Nucl. Phys.} {\bf
B544}  (1999) 243;\
M. Ro\~cek and A. Tseytlin, { Phys. Rev.} {\bf D59} (1999)
106001;\  E. Ivanov and S. Krivonos,
{ Phys. Lett.\/} {\bf B453} (1999) 237;\
S. V. Ketov, { Mod. Phys. Lett.\/} {\bf A14} (1999) 501;\
{ Born-Infeld-Goldstone superfield actions for gauge-fixed D-5- and
D-3-branes in 6d\/}, hep-th/9812051.
\item{14}
T. Adawi, M. Cederwall, U. Gran, M. Holm and B. E. W. Nilsson,
{ Int. J. Mod. Phys.\/} {\bf A13} (1998) 4691.
\item{15} O. Barwald and P. West, { Brane Rotating symmetries and
the fivebrane equations of motion} hep-th/9912226.
\item{16} E. A. Ivanov and V. I. Ogievetsky, { Teor. Mat. Fiz.\/}
{\bf 25} (1975) 164. 
\item{17}
P. S. Howe and  E. Sezgin, {  Phys. Lett.} B390 (1997) 133;
P. S. Howe and  E. Sezgin, { Phys. Lett.} B394 (1997) 62.
\item{18}
D. Sorokin and P. K. Townsend.
{ M theory superalgebra from the M five-brane.};
{ Phys. Lett.}, B412 (1997) 265, hep-th/9708003.
\item{19}
T.  Adawi, M. Cederwall, U. Gran, B. E. W. Nilsson and B. Razaznejad,
JHEP {\bf 9902} (1999) 001.
\item {20}
E. Ivanov, S. Krivonos, N=1 D=4 supermembrane in the coset approach
 Phys.Lett. B453 (1999) 237, hep-th/9901003.
\item{21} F. Delduc, E. Ivanov, S. Krivonos,
 1/4 PBGS and Superparticle Actions, hep-th/9912292;
F Delduc, E. Ivanov, S. Krivonos, 1/4 Partial Breaking of
Global Supersymmetry and New Superparticle Actions,  hep-th/9912222
\item{22} Igor Bandos, Jerzy Lukierski, Dmitri Sorokin,
The OSp(1/4) Superparticle and Exotic BPS States, hep-th/9912264
; Paolo Pasti, Dmitri Sorokin, Mario Tonin,
  Branes in Super-AdS Backgrounds and Superconformal Theories,
hep-th/9912076
; Igor Bandos, Kharkov, Ukraine, Jerzy Lukiersk, Dmitri Sorokin,
Generalized Superconformal Symmetries and Supertwistor Dynamimics,
hep-th/9912051
\item{23}
J. A. de Azcarraga, J. P. Gauntlett, J. M. Izquierdo
and P. K. Townsend; Topological extensions of the superalgebra for
extended objects; { Phys. Rev. Lett.}, {\bf B63} (1989) 2443.
\item{24} J. P. Gauntlett, K,Itoh and P. K. Townsend, Superparticle 
with extrinsic curvature, Phys. Lett.  B (1990) 65; 
J. P. Gauntlett, J. Gomis and P. K. Townsend, Particle actions as Wess-Zumino 
terms for space-time (super)symmetry groups, Phys. Lett.  B (1990) 255.

\end